\documentclass[conference, a4paper, final]{IEEEtran}
\IEEEoverridecommandlockouts
\usepackage{cite}
\usepackage{amsmath,amssymb,amsfonts}
\usepackage{algorithmic}
\usepackage{graphicx}
\usepackage{textcomp}
\usepackage{xcolor}
\def\BibTeX{{\rm B\kern-.05em{\sc i\kern-.025em b}\kern-.08em
    T\kern-.1667em\lower.7ex\hbox{E}\kern-.125emX}}
\begin{document}

\title{Modelling for Quantum Error Mitigation\\

\thanks{We acknowledge the support by DASHH (Data Science in Hamburg - HELMHOLTZ Graduate School for the Structure of Matter) with the Grant-No. HIDSS-0002.}
}

\author{\IEEEauthorblockN{Tom Weber\IEEEauthorrefmark{1}, Matthias Riebisch\IEEEauthorrefmark{1}, Kerstin Borras\IEEEauthorrefmark{2}\IEEEauthorrefmark{3}, Karl Jansen\IEEEauthorrefmark{2}\IEEEauthorrefmark{4} and Dirk Krücker\IEEEauthorrefmark{2}}\\
\IEEEauthorblockA{\IEEEauthorrefmark{1}
Universität Hamburg\\
Hamburg, Germany\\
Email: tweber@informatik.uni-hamburg.de}\\
\IEEEauthorblockA{\IEEEauthorrefmark{2}Deutsches Elektronen-Synchrotron DESY\\
Hamburg, Germany}\\
\IEEEauthorblockA{\IEEEauthorrefmark{3}RWTH Aachen University
\\
Aachen, Germany}\\
\IEEEauthorblockA{\IEEEauthorrefmark{4}John von Neumann Institute for Computing\\
Zeuthen, Germany}}
\maketitle

\begin{abstract}
While we expect quantum computers to surpass their classical counterparts in the future, current devices are prone to high error rates and techniques to minimise the impact of these errors are indispensable. There already exists a variety of error mitigation methods addressing this quantum noise that differ in effectiveness, and scalability. But for a more systematic and comprehensible approach we propose the introduction of modelling, in particular for representing cause-effect relations as well as for evaluating methods or combinations thereof with respect to a selection of relevant criteria.
\end{abstract}


\section{Introduction}
Quantum computers are widely believed to outperform classical machines in a variety of tasks and applications in the future \cite{b1,b2,b3,b4,b5,b30}. However, current devices are prone to errors, yielding noisy computational results \cite{b6}. Various causes of this noise have been identified ranging from faulty qubit control including the preparation, manipulation, and measurement of quantum states, to undesired interaction of qubits with each other or their environment. Moreover, the impact of errors grows with the number of qubits involved and the number of gate operations applied in a quantum algorithm, making quantum computation impractical in many situations. Although there is a prospective solution to this problem in the form of fault-tolerant quantum computation, it is currently still unavailable due to small qubit resources and high error rates \cite{b7,b8}. If the rates were below a certain threshold, we could apply error correction codes to effectively address noise by encoding quantum information in logical qubits composed of several physical ones \cite{b9,b10}. But until fault-tolerance is achieved other methods are required in order to make use of quantum computers in the near future, at least for specific problems. This gives rise to the following question: How can we deal with the flaws of quantum devices?\par

Answering this question is the goal of quantum error mitigation which has been gaining much attention in recent years. By now there exists a variety of mitigation schemes tackling different types of noise within different contexts of application \cite{b14,b15,b16,b17,b19}. Among the most promising candidates for the latter ones are hybrid algorithms that comprise both classical and quantum components \cite{b11,b12,b13}. Typically, the majority of the algorithm is carried out on a classical machine and only a fraction of the program is executed on a quantum computer.\par

All methods aim to derive, or at least approximate, error-free computational results from faulty ones by taking into account the effect of noise and many of them are designed for the context of hybrid algorithms. Especially measurement errors and imperfect gate operations have been widely investigated and various ideas have been put forward to model these errors, construct calibration procedures in order to obtain relevant model parameters and achieve an improvement of results. For instance, in \cite{b20} the authors propose a way to mitigate readout error in the context of computing expectation values of observables by modifying the measured operators accordingly. Moreover, it has been suggested to use machine learning in quantum error mitigation \cite{b21,b22}, resulting in a very diverse field of research.\par 

What is missing in this diversity is systematisation. Even though many mitigation methods are constructed for the same task, there is no common metric to evaluate them with respect to factors such as effectiveness, scalability and computational effort and although these are discussed to some extent in most of the work, there is effectively no comparability. Furthermore, the combination of different procedures is rarely considered\cite{b23}.\par

In this paper we propose the novel approach to introduce modelling concepts and techniques from computer science to quantum error mitigation. In general, models are simplified representations of systems that can help us understand, predict, and therefore possibly control their complex correlations \cite{b24,b25,b26}. They can also be used for communication and reduce information to the essentials with respect to a specific point of view. We present several possible views on quantum error mitigation where we want to integrate models, including cause-effect relationships of quantum noise as well as costs and benefits of mitigation schemes and combinations thereof. \par

There are many disciplines which have successfully incorporated modelling techniques and even in quantum computing we find a multitude of examples where models are already being used. The most prominent one might be the circuit model\cite{b30} which represents algorithms by quantum circuits analogous to logical circuits in classical computation. Similarly, there are other models that concern the general workings of quantum computers. First attempts have been made towards a framework for modelling quantum software by extending the Unified Modeling Language (UML) \cite{b28}. But none of this work directly aims at quantum error mitigation and since quantum noise is the main obstacle in our way to practically benefit from quantum computing, we expect a systematic approach to be helpful.\par

The rest of this work is structured as follows: In section \ref{sec:modelsinquantumerrormitigation} we present our ideas of how to integrate modelling and in the final section \ref{sec:summaryandoutlook} we summarise the suggestions and give an outlook on possible future work in this direction.

\section{Models in Quantum Error Mitigation}
\label{sec:modelsinquantumerrormitigation}
In this section we discuss in more depth the types of models we propose as well as the applications for which we anticipate them to be useful. We first consider the complex cause-effect relationships of quantum noise and outline an attempt to model them with a fishbone diagram. Afterwards we examine the evaluation of different mitigation techniques with the goal to create more comparability. In the end we give an outlook on how to generalise this to combinations of several methods.
\subsection{Cause-effect relationships of quantum noise}

In order to effectively and efficiently mitigate errors, we need an exact understanding of the responsible processes and their influence on each other. Depending on the computational task to be performed it is necessary to take into account different aspects of the system. To demonstrate this, we will look at variational quantum eigensolvers (VQEs) \cite{b12} which are hybrid classical/quantum algorithms consisting of a classical optimiser that minimises a cost function evaluated on a quantum computer. This example is not only relevant for applications in the near future, but it is also instructive for the purpose of this work. \par

\begin{figure}[b]
\includegraphics[width=0.5\textwidth]{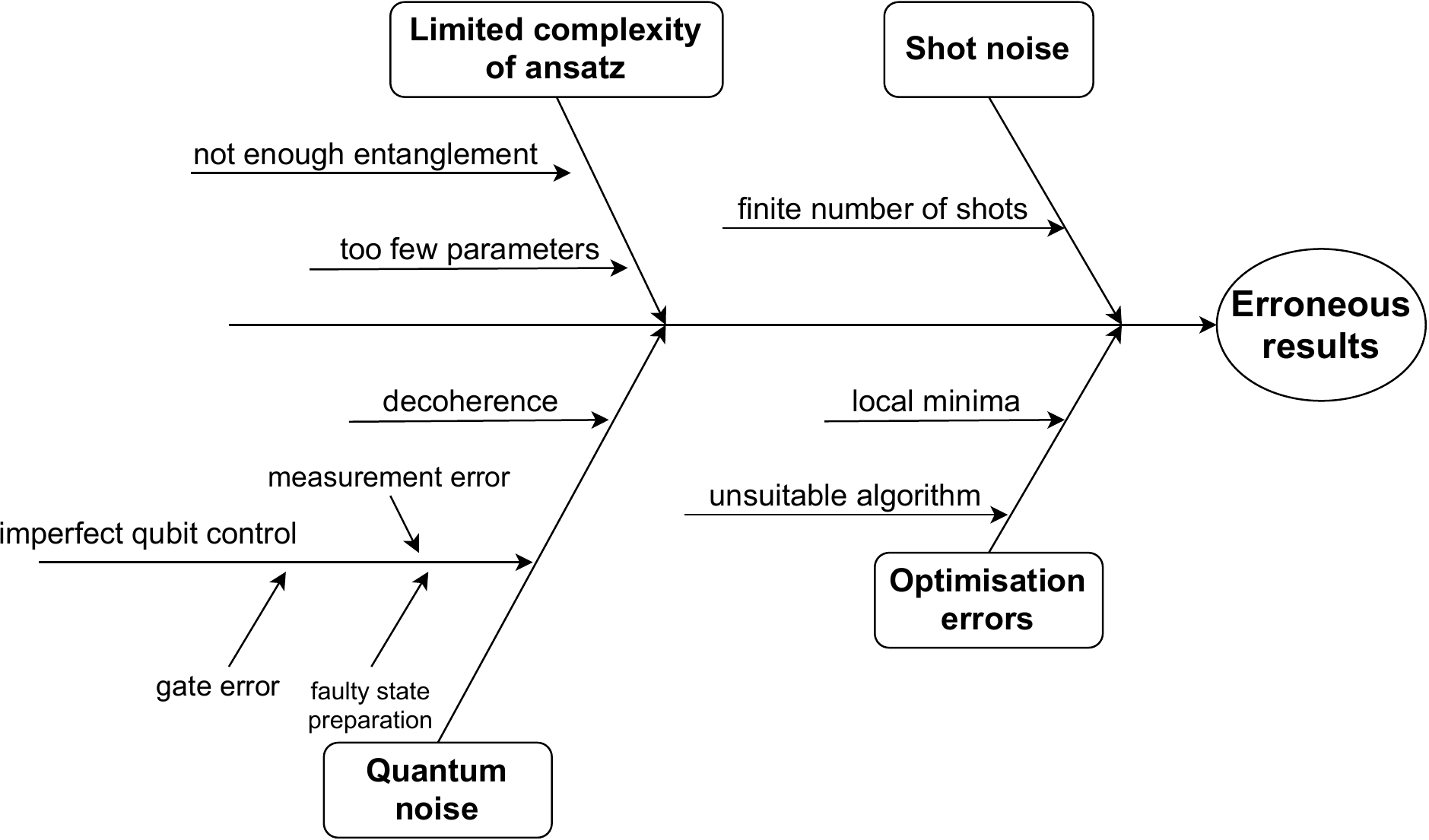}
\caption{Example of an Ishikawa diagram for modelling cause-effect relationships in error mitigation.}
\label{fig}
\end{figure}

The modelling of cause-effect relationships has been investigated in a variety of research fields and is often centred around Ishikawa diagrams \cite{b29}, also known as fishbone diagrams. This is the first model we want to discuss, so we start with a brief explanation of its structure. Given a certain effect one identifies its causes and iteratively repeats this procedure for each of the causes, yielding a hierarchical structure which can be depicted graphically as a fishbone. Figure \ref{fig} shows an exemplary Ishikawa diagram for VQEs. Note that this is not meant to be complete, but we rather want to give a short outlook.\par

The diagram gives a comprehensive overview of the processes that impact our computations and helps us structure our knowledge. For instance, we see that there are far more sources of noise than just quantum errors (see e.g. \cite{b31} for the complexity). Moreover, the tree structure categorises the processes and therefore organises the information. But it also suggests that different branches are independent of each other which is not necessarily true. Furthermore, causes with almost negligible impact are represented in the same manner as the most relevant ones. Thus, in order to make this model more useful, modifications are necessary. In the following we present two main features to be added.\par

In \cite{b27} we find suggestions to modify Ishikawa diagrams in terms of more logic being depicted, for example necessary or sufficient conditions of causes and effects. Our first idea is to integrate this into the context of error mitigation, giving the graphical structure a formal meaning and enabling us to represent the processes and their consequences more precisely. The second extension we propose is to include different degrees of relevance of the branches in the model. This could be done by additional labels indicating how strong a certain cause impacts the computation.\par

With these two additions the model can help us not only to structure our knowledge, but also manage the complexity of quantum errors by representing the corresponding sources of error and their interactions as precisely as possible. Moreover, we can benefit from this model when it comes to prioritisation of different types of noise, clarifying which causes should primarily be focussed to improve results.

\subsection{Evaluating mitigation methods}
The numerous mitigation methods that have been developed in recent times deviate from each other with respect to the specific task they can be applied to, the types of error they address, their impact on computational results, and other properties which we will discuss later. If two mitigation schemes are designed for the same task, the question arises: Which one should be used? Previous work often has been limited to benchmarking new methods in seemingly arbitrary test settings and comparing them to only a few selected other routines with respect to varying aspects. We outline a systematic approach for the comparison of quantum error mitigation schemes to support decision making for quantum software.
\par

For a meaningful evaluation of mitigation techniques it is necessary to choose sensible criteria. The relevance thereof depends on the context, but there are some features which will be important independent of the specific application. Before we introduce these necessary attributes, let us clarify what our assumptions are. Since comparing mitigation methods only makes sense if they aim at the same type of computation, we limit ourselves to this case. Moreover, we assume that all prerequisites of the methods can be met. Either they are fulfilled a priori or this can be achieved by additional effort which then must be taken into account by evaluation of the corresponding feature. Our criteria are:

\begin{itemize}
\item \textbf{Degree of improvement}: As quantum error mitigation always aims to minimise the deviation of quantum computational results from the correct ones, the degree of error reduction obviously has to be considered. We suggest constructing common test experiments reflecting the nature of the target application and compare the results. Since the outcome of our tests should be as expressive as possible, it makes sense to have a quantitative rating whenever it is possible.\par

For instance, let us again consider the context of VQEs where the quantum device computes a cost function. This function is typically the expectation value of a quantum mechanical observable which we can calculate exactly with a classical machine (at least for relatively small qubit numbers).  So we could create a large set of benchmark experiments, compute the relative errors of our results and take the mean and standard deviation as an evaluation outcome. Note that the standard deviation can give valuable information about the reliability of the methods.

\item \textbf{Computational effort (classical/quantum)}: Every benefit we gain from error mitigation comes with a certain cost. Often there are additional routines to be run both on classical and on quantum devices. These can include training of machine learning models, running calibration experiments to obtain relevant parameters, and many more. If we want to incorporate any mitigation technique, additional computational resources are needed and our evaluation should represent this fact. Unfortunately, it is hard to explicitly quantify this cost, so in most cases we can only make a qualitative assessment.

\item \textbf{Scalability}: The additional effort discussed above depends on the number of qubits and gate operations that are involved. Since this dependency is crucial for the practicability, we list this feature separately from the effort itself. Even the mitigation scheme with the lowest additional effort for low qubit numbers can become useless for larger systems if the overhead grows exponentially.
\end{itemize}

Leaving any of these three features out of consideration makes a useful evaluation impossible because for each of the feature evaluations there are outcomes that make a mitigation method entirely impractical. For instance, if the additional resources needed are higher than the possibilities of a quantum computing practitioner, she cannot incorporate the method at all. In that sense the degree of improvement, computational cost and scalability are necessary evaluation criteria for quantum error mitigation. Note that we do not claim this list to be complete; depending on the context there might be more attributes that cannot be neglected. \par

In order to actually support decisions about the architecture of quantum programs, good models for the single features as well as the overall evaluation are necessary. In the following we want to consider a simple example to justify this statement. Assume we can assign a meaningful rating score to each attribute as well as weights for the purpose of prioritisation. A straightforward way to generate a total evaluation score is to compute the weighted sum. In principle this makes it possible to compare different mitigation methods but only if the scores of the single criteria represent the behaviour of the mitigation technique well enough, will our evaluation be of any use. Since the reduction of information to the essentials is one of the main strengths of modelling, we expect to benefit from their introduction in this context.

\subsection{Combining methods}
Most of the existing error mitigation techniques tackle a certain type of error. When there are several methods applicable for a task and they address the same problem, one could choose between them. But if they counteract errors which are considered to be independent of each other (e.g. readout and gate error), it might even be possible to combine them. In the first scenario the choice leads us back to the discussion about evaluating mitigation schemes above, whereas for the second scenario the question arises: How to do it?\par

To explain this question, consider the situation where we want to compute a quantum mechanical expectation value with a quantum computer. In \cite{b20} and \cite{b32} we find methods to tackle measurement error and depolarisation for this task, respectively. Both demand the execution of certain calibration experiments to obtain parameters of the corresponding error model. After we have computed these parameters we can run the original experiment and correct the results accordingly. But whether we should calibrate simultaneously or in a certain order (already correcting the calibration results of one method with the other one) remains unclear. The same holds for the different post-processing steps of the methods.\par

If we can get a good understanding on how different mitigation methods affect each other, we can potentially find the best way to combine them and thereby tackle several sources of noise at once. This brings us back to modelling cause-effect relationships. With a good model in this setting one could establish a general procedure to create effective combinations of methods which then can be evaluated with a metric as proposed in the section above. Summarising, progress in the two previous fields will also improve our comprehension in this direction.

\section{Summary and outlook}
\label{sec:summaryandoutlook}
In this paper we have proposed the explicit incorporation of modelling techniques into quantum error mitigation. We outlined possible benefits for better comprehension of noise and support of decision making for near-future quantum software architecture. The specific applications that we have put forward are cause-effect models for the noise itself and for constructing combinations of mitigation methods as well as the evaluation thereof. We exemplified these aspects with simple models and discussed their advantages and limitations.\par

For the cause-effect models we have observed that the widespread Ishikawa diagram already can help us structure our knowledge about quantum noise and therefore improve our comprehension but it cannot represent the interactions of errors in sufficient detail. Hence, extensions of this model are needed and we have presented two important features that we want to add in the future. The fact that our examples could already bring some degree of benefit despite their simplicity suggests that models with more depth can go beyond this and actually advance progress.\par

Concerning the evaluation of mitigation methods, we have advertised the use of common criteria and stated three attributes which should always be taken into account when appraising a routine. Moreover, we have argued that good models for these criteria can lead to the construction of  meaningful metrics and therefore to new standards for the comparison  of mitigation methods.\par

In the future the simple ideas from above will be worked out in more detail and with more refined models, in particular for the assessment criteria. In the best case there will be an agreement on how to evaluate quantum error mitigation methods, but even if not we will still end up with a guideline helping to choose between different schemes and understanding the impact of those. Moreover, there are other potential benefits from modelling which we did not yet consider. For instance, models can be used as a tool for communication. Since quantum computing may become interesting for many different fields (e.g. computer science, chemistry, physics), they can provide a common language between different participants.

\end{document}